\documentclass[12pt]{article}
\usepackage{amsthm,color}
\usepackage{amsfonts}
\usepackage{graphicx}
\usepackage{mathrsfs}
\usepackage{amsmath}
\usepackage{amssymb}
\usepackage{float}
\usepackage{natbib}
\usepackage{booktabs}
\usepackage{url}
\usepackage{multirow}
\usepackage{arydshln}
\usepackage{threeparttable}
\usepackage{courier}
\usepackage{authblk} 
\usepackage{moreverb}
\usepackage{multicol}
\usepackage{latexsym}
\usepackage{psfrag}
\usepackage[usenames,dvipsnames]{xcolor}
\usepackage{enumitem}
\usepackage{bm}
\usepackage{breakcites}
\usepackage{bbm}
\usepackage{epsfig,epstopdf}
\usepackage[colorlinks, linkcolor=black, citecolor=black]{hyperref}

\usepackage[utf8]{inputenc}
\usepackage[english]{babel}
\usepackage{caption}
\usepackage{setspace} 

\usepackage{natbib} 
\usepackage{sectsty}
\sectionfont{\fontsize{12}{15}\selectfont}
\subsectionfont{\fontsize{12}{15}\selectfont}

\newtheorem{theorem}{Theorem}
\newtheorem{proposition}{Proposition}

\def\ba{\left( \begin{array}}
\def\ea{\end{array} \right)}

\begin{document}

{\centering {\large {\bf Receiver operating characteristic curve analysis  with non-ignorable missing disease status}}\par }
\bigskip

\centerline{Dingding Hu, Tao Yu, and \ Pengfei Li\footnote{Dingding hu is a doctoral student, Tao Yu is Associate Professor, Department of Statistics \& Applied Probability, National University of Singapore, Singapore, Pengfei Li is Professor, Department of Statistics and Actuarial Science, University of Waterloo, Waterloo ON N2L 3G1,Canada (E-mails:{\em d32hu@uwaterloo.ca},  \ {\em yu.tao@nus.edu.sg} \ and \  {\em pengfei.li@uwaterloo.ca}).}}

\bigskip

\bigskip

\hrule

{\small
\begin{quotation}
\noindent
This article considers the receiver operating characteristic (ROC) curve analysis for medical data with non-ignorable missingness in the disease status. In the framework of the logistic regression models for both the disease status and the verification status, we first establish the identifiability of model parameters, and then propose a  likelihood method to estimate the model parameters, the ROC curve, and the area under the ROC curve (AUC) for the biomarker. The asymptotic distributions of these estimators are established. Via extensive simulation studies, we compare our method with competing methods in the point estimation and assess the accuracy of confidence interval estimation under various scenarios.
To illustrate the application of the proposed method in practical data, we apply our method to the National Alzheimer's Coordinating Center data set.
\vspace{0.3cm}

\end{quotation}
}

\hrule

\bigskip

\bigskip

\section{Introduction}

In medical research, biomarkers are commonly used to differentiate between healthy and diseased patients. According to studies by \cite{Yin2014} and \cite{Hu2023}, it is assumed that biomarker values are higher in diseased individuals than in healthy ones. Under this assumption, an individual is classified as diseased if their biomarker value exceeds a certain threshold $x$. Sensitivity and specificity are introduced as measurements of how well such a test can identify whether the individual is diseased or not. Specifically, sensitivity and specificity are defined to be $1-F_1(x)$ and $F_0(x)$, where $F_0(\cdot)$ and $F_1(\cdot)$ are respectively the cumulative distribution functions (cdfs) of the biomarker for healthy and diseased populations. In practice, it is important to select a reasonable $x$, such that these measurements can have a well trade-off, and thus give reasonable testing results \citep{Zhou2002}. The receiver operating characteristics (ROC) curve illustrates the relationship between these measures by plotting sensitivity ($1-F_1(x)$) against one minus specificity ($1-F_0(x)$), making it a widely used statistical tool for evaluating the performance of a binary classifier across various discrimination thresholds $x$ \citep{Zhou2002,Yuan2021}. Mathematically, the ROC curve is given by 
\begin{equation}
\label{def.roc}
ROC(s)=1-F_1\big(F_0^{-1}(1-s)\big),
\end{equation}
for $s\in(0,1)$, 
where 
$
    F^{-1}_0(1-s)=\inf\left\{x:F_0(x)\geq 1-s\right\}.
    $
A commonly used summary index of the ROC curve is the area under the curve (AUC), which is defined as:
\begin{equation}\label{def.auc}
    AUC=\int_{0}^1ROC(s)ds    =\int F_0(x) dF_1(x).
\end{equation}

The classical ROC curve estimation methods typically require the knowledge of the ``true disease status" of all the patients in the study, which are usually determined by a gold standard test. However, in many applications, verifying disease status for every individual is impractical due to cost, time, or ethical constraints, leading to data with missing disease status for some patients. For these data, a popular assumption is that the missing disease status is missing at random (MAR): the verification status is independent of the true disease status, given the available biomarker and covariate information. Under this assumption, the ROC estimation has been well established in the literature, such as  \cite{Begg1983}, \cite{Zhou1998}, \cite{Alonzo2009}, \cite{Zhou2002}, and the references therein.

However, the MAR assumption is restrictive and is often violated in practice. For example, consider the Alzheimer’s disease (AD) study discussed in \cite{Zhou2004}. In that study, a two-stage design is employed, where the first stage involves a screening test on a large population, and the second consists of a more comprehensive and costly clinical diagnosis on a selected subgroup of individuals from the first stage. The selected patients in the subgroup for the second stage depend on the test results from the first stage and other relevant covariates. The collected data from such a study do not satisfy the MAR assumption, since the diagnotic test in the second stage is applied only to individuals selected based on a mechanism that may relate to their disease status. With this data, the statistical inference for the ROC curve and AUC, based on the MAR assumption, can be biased \citep{Liu2010}.

Assuming the existence of the non-ignorable verification bias, the estimation method for the AUC is available in the literature. \cite{Liu2010} proposed a method to model such data with two separate logistic regression models: one regress the disease status on the biomarker and relevant covariates, and the other regress verification status on the disease status, biomarker, and relevant covariates. However, as noted by \cite{Liu2010}, it may not be feasible to validate these models because of the missing disease status in data.  To overcome this difficulty, \cite{Yu2018} proposed regressing the disease status on the biomarker and relevant covariates only for individuals with verified disease status, while retaining the same model for verification status as in \cite{Liu2010}.


We note that \cite{Yu2018}'s method has three limitations. First, their estimation method for the unknown parameters does not make full use of the available data, leaving potential for improvement. Their AUC estimator relies on the inverse probability weighting (IPW) method, which uses data only from individuals with verified disease status, potentially making it less efficient than methods that utilize all available data \citep{Liu2010, Liu2022}. Second, \cite{Yu2018} assumes the availability of an instrumental variable; in practice, however, it may not be clear of how to identify it. Third, statistical inference for the ROC curve is not considered.
 
Our contribution in this paper is threefold. First, under the same framework as \cite{Yu2018}, we establish the identifiability of model parameters under mild conditions, without requiring the existence of an instrumental variable. Second, we propose a maximum likelihood method to estimate the unknown model parameters, ROC curve, and AUC. Our approach incorporates all available data, and is therefore expected to be more efficient than the method in \cite{Yu2018}; this has been demonstrated in our simulation study. Third, we introduce a two-step verification process for both the disease and verification models.

The rest of this paper is organized as follows. In Section \ref{Chapter4-estimation}, we introduce our proposed method, detailing the estimation of the ROC curve and AUC, model identifiability, and model verification. The asymptotic properties of the ROC curve and AUC estimators are discussed in Section \ref{chapter4-asymptotic}. In Section \ref{chapter4-simulation}, we present simulation studies to compare the performance of our method with existing methods. Section \ref{realdata} illustrates our proposed method using a real data example, and the paper concludes in Section \ref{conclude}.

\section{Proposed method}
\label{Chapter4-estimation}

\subsection{Model setup}
\label{setup}
For an individual, let $Y$ denote their disease status (0 = healthy, 1 = diseased), $X$ their biomarker, and $\bm{V}$ their associated covariates. Let $R$ represent the verification status or missing indicator for $Y$, where $R = 0$ indicates that $Y$ is missing, and $R = 1$ indicates that $Y$ is observed.
In the framework of \cite{Yu2018}, we consider the disease model:
\begin{eqnarray} 
P(Y=1|x,{\bm v},R=1) &\equiv & P(Y=1|X=x,{\bm V}={\bm v},R=1)\nonumber\\
&=&\frac{1}{1+\exp(\mu_1+\mu_2x+{\bm\mu}_3^T{\bm v})}; \label{disease-model}  
\end{eqnarray}
and the verification model:
\begin{eqnarray}
\nonumber P(R=1|y,x,{\bm v}) &\equiv & P(R=1|Y=y,X=x,{\bm V}={\bm v})\\
&=&\frac{1}{1+\exp(\psi_1+\psi_2 x+{\bm\psi}_3^T{\bm v}+\beta y)}.\label{verification-model} 
\end{eqnarray}

Let $F_1(x) = P(X \leq x | Y = 1)$ and $F_0(x) = P(X \leq x | Y = 0)$.  In this paper, we focus on statistical inference for the ROC curve and AUC under the models specified above, where ROC curve and AUC are given by \eqref{def.roc} and \eqref{def.auc}. For presentational convenience, we denote 
${\bm\mu}=(\mu_1,\mu_2,{\bm\mu}_3^T)^T$,  ${\bm\phi}=(\psi_1,\psi_2,{\bm\psi}_3^T,\beta)^T$, and $P_{1}(x,\bm{v};{\bm\mu})=P(Y=1|x,\bm{v},R=1)$.

\subsection{Identifiability of model parameters}\label{model-identification}
We consider the identifiability and the needed conditions for models \eqref{disease-model} and \eqref{verification-model}. For individuals with $R=1$, the corresponding $Y, X, \bm{V}$ are all observed; the model \eqref{disease-model} is essentially the classical logistic regression model. Thus, its model parameters $\bm{\mu}$ are identifiable. 



We proceed to consider the identifiability of ${\bm\phi}$ in \eqref{verification-model}. We have
\begin{align}\label{ratio1}
    \frac{P(X=x, {\bm V}={\bm v}, Y=y, R=0)}{P(X=x, {\bm V}={\bm v}, Y=y, R=1)} & = \exp(\psi_1+\psi_2 x+{\bm\psi}_3^T{\bm v}+\beta y),
\end{align}
which leads to 
\begin{align*}
    P(X=x, {\bm V}={\bm v}, R=0) & = \sum_{y=0}^1 P(X=x, {\bm V}={\bm v}, Y=y, R=0 )\\
    & = P(X=x, {\bm V}={\bm v}, R=1) \exp \{\psi_1+\psi_2 x+{\bm\psi}_3^T{\bm v}+c(x,{\bm v};{\bm \mu},\beta)\},
\end{align*}
where 
\begin{equation}
\label{formc.ch4}
c(x,{\bm v};{\bm \mu},\beta)=\log E(e^{\beta Y}|x,{\bm v}, R=1)
=\log\left\{\exp(\beta)P_{1}(x,\bm{v};{\bm\mu})
+1-P_{1}(x,\bm{v};{\bm\mu})\right\}.
\end{equation}
Applying Bayes' formula, 
we have 
\begin{eqnarray}
P(R=1|x, {\bm v}) &\equiv & P(R=1|X=x, {\bm V}={\bm v})
\nonumber\\
&
=&\frac{1}{1+\exp \{\psi_1+\psi_2 x+{\bm\psi}_3^T{\bm v}+ c(x,{\bm v};{\bm \mu},\beta)\}}.
\label{drm.ch4}
\end{eqnarray}
Because $(R, X, \bm{V})$ are observed for all the individuals, $P(R = 1 | x, \bm{v})$ is identifiable. As a consequence, based on \eqref{drm.ch4}, the identifiability of $\bm{\phi}$ is equivalent to its identifiability in the expression $\psi_1 + \psi_2 x + \bm{\psi}_3^T \bm{v} + c(x, \bm{v}; \bm{\mu}, \beta)$. Theorem \ref{theorem-identifiability} below ensures the identifiability of the parameters under mild regularity conditions. 

\begin{theorem}\label{theorem-identifiability}
Suppose Condition C1 in the Appendix is satisfied. Then, the model parameters 
${\bm\mu}$ in \eqref{disease-model} and 
${\bm\phi}$ in 
\eqref{verification-model} are identifiable.
\end{theorem}

For presentational continuity, we put the technical details of this theorem in Section 1 of the supplementary material. We emphasize that, based on Theorem \ref{theorem-identifiability}, the identifiability of $({\bm \mu}, {\bm \phi})$ does not depend on the existence of an instrumental variable. Therefore, in practice, our method can be applied without the need to identify such a variable.

\subsection{Estimation of model parameters}
\label{Parameter}
 
Let $\{ (Y_i, X_i,{\bm V_i},R_i), i=1, \ldots, n\}$ be independent and identically distributed copies of $(Y, X, {\bm V},R)$, where the biomarker $X_i$ and the covariates ${\bm V_i}$ are always observed,
and $Y_i$ is observed if and only if $R_i=1$. 
Without loss of generality, we assume that $R_i=1$,  for $i=1,\ldots,n_1$,
and $R_i=0$, for $i=n_1+1,\ldots,n$.

The likelihood for the observed  data is given by 
\begin{align*}
    L_n& =\prod_{i=1}^{n_1}P(Y_i,R_i=1|X_i,{\bm V}_i)\cdot \prod_{i=n_1+1}^{n}P(R_i=0|X_i,{\bm V}_i)\cdot \prod_{i=1}^n P(X_i,{\bm V}_i)\\
    & = \prod_{i=1}^{n_1}P(Y_i|X_i,{\bm V}_i, R_i=1)\cdot  \prod_{i=1}^n P(R_i|X_i,{\bm V}_i)\cdot \prod_{i=1}^n P(X_i,{\bm V}_i).
\end{align*}
Thus, after ignoring a constant that does not depend on $\bm\mu$ and $\bm\phi$, the log-likelihood is given by 
$$
l_n({\bm\mu},{\bm\phi})
=l_{n1}(\bm\mu)+l_{n2}({\bm\mu},{\bm \phi}), 
$$
where 
$$
l_{n1}(\bm\mu)=
\sum_{i=1}^n R_i\left[Y_i\log \left\{P_{1}(X_i,\bm{V}_i;\bm{\mu})\right\}+(1-Y_i)\log \left\{1-P_{1}(X_i,\bm{V}_i;\bm{\mu})\right\}\right]
$$
and 
$$
l_{n2}({\bm\mu},{\bm \phi})=\sum_{i=1}^n
[
R_i\log \pi(X_i,{\bm V}_i;{\bm\mu},{\bm \phi})
+(1-R_i)
\log\{1- \pi(X_i,{\bm V}_i;{\bm\mu},{\bm \phi}) \}
],
$$
where $P_{1}(x,\bm{v};{\bm\mu})=P(Y=1|x,\bm{v},R=1)$ is given 
in \eqref{disease-model}
and $\pi(x,{\bm v};{\bm\mu},{\bm \phi})=P(R=1|x, {\bm v})$ is given in \eqref{drm.ch4}. 
The maximum likelihood estimator of $({\bm\mu},{\bm\phi})$ is then defined as
\begin{equation}\label{parameter-estimte}
(\hat{\bm\mu},\hat{\bm\phi})=\arg \max_{{\bm\mu},~{\bm\phi}} l_n({\bm\mu},{\bm\phi}).
\end{equation}

\subsection{Estimation of ROC curve and AUC\label{ROC}}
Based on the parameter estimators, we proceed to establish estimators for $F_0(\cdot)$ and $F_1(\cdot)$, and then use them to establish estimators for the ROC curve and AUC for the biomarker. 
Note \eqref{verification-model} and \eqref{drm.ch4}. Applying Bayes's rule, we have 
\begin{eqnarray}
\nonumber
\frac{P(Y=1|x,{\bm v},R=0)}{P(Y=1|x,{\bm v},R=1)}&=&\frac{P(R=0|x,{\bm v},Y=1)/P(R=0|x,{\bm v})}{P(R=1|x,{\bm v},Y=1)/P(R=1|x,{\bm v})}\\
&=&\frac{\exp(\beta)}
{\exp\{c(x,{\bm v};{\bm \mu},\beta)\}},
\nonumber
\end{eqnarray}
which, together with model \eqref{disease-model}, leads to 
\begin{align*}   
P(Y=1|x,{\bm v},R=0)&=\frac{\exp(\beta)}
{\exp\{c(x,{\bm v};{\bm \mu},\beta)\}} 
P(Y=1|x,{\bm v},R=1)\\
&=\frac{\exp(\beta)P(Y=1 | x,{\bm v},R=1)}{1-P(Y=1 | x,{\bm v},R=1)+P(Y=1 | x,{\bm v},R=1)e^{\beta}}\\
&=\frac{1}{1+\exp(\mu_1+\mu_2x+{\bm\mu}_3^T {\bm v}-\beta)}. 
\end{align*}
Referring to model \eqref{disease-model}, we can write 
\begin{eqnarray}
P(Y=1|x,{\bm v},r)  & \equiv & P(Y=1|X=x,{\bm V}={\bm v},R=r)
\nonumber\\
&=&\frac{1}{1+\exp\{ \mu_1+\mu_2 x+{\bm\mu}_3^T {\bm v}+(r-1)\beta\}}.\label{def.gfun}
\end{eqnarray} 
Denote $g(x,{\bm v},r;{\bm\mu},\beta) =P(Y=1|x,{\bm v}, r)$. The following proposition establishes the connection between $F_0(\cdot)$, $F_1(\cdot)$ and $g(x,{\bm v},r;{\bm\mu},\beta)$. The proof is given in Section 2 of the supplementary material. 

\begin{proposition}\label{proposition1}
For any $x$, we have
$$
E\left\{
g(X,V,R;{\bm\mu},\beta)I(X\leq x)
\right\}
=E\left\{
g(X,V,R;{\bm\mu},\beta)
\right\}\cdot F_1(x)
$$
and 
$$
E\left[
\{1-g(X,V,R;{\bm\mu},\beta)\}I(X\leq x)
\right]
=E\left[\left\{
1-g(X,V,R;{\bm\mu},\beta)
\right\}\right]\cdot F_0(x).
$$
\end{proposition}
Based on Proposition \ref{proposition1}, we can estimate $F_0$ and $F_1$ by
\begin{equation*}
\hat{F}_1(x)
=
\frac{\sum_{i=1}^n \hat {g}_i I(X_i\leq x) }{\sum_{i=1}^n \hat {g}_i}
\end{equation*}
and
\begin{equation*}
\hat{F}_0(x)=\frac{\sum_{i=1}^n (1-\hat {g}_i) I(X_i\leq x) }{\sum_{i=1}^n (1-\hat {g}_i)},
\end{equation*}
where $ \hat {g}_i=g(X_i,{\bm V}_i,R_i;\hat{\bm\mu},\hat\beta).$
Consequently, the ROC curve and AUC are estimated by 
\begin{equation*}
    \widehat{ROC} (s)= 1-\hat F_1\big(\hat F_0^{-1}(1-s)\big )
\end{equation*}
and 
\begin{equation*}
    \widehat{AUC}=
    \int \hat F_0(x)d\hat F_1(x).
\end{equation*}
The asymptotic properties of $\widehat{ROC} (s)$ and $\widehat{AUC}$ will be studied in 
Section \ref{chapter4-asymptotic}.

\subsection{Model verification}
\label{model verification}
The estimation procedures outlined above rely on the model assumptions specified in \eqref{disease-model} and \eqref{verification-model}. This section describes how to verify these assumptions using the available data $\left\{ (Y_i, X_i, \bm{V}_i, R_i), i = 1, \dots, n \right\}$. We verify models \eqref{disease-model} and \eqref{verification-model} separately. 

Since model \eqref{disease-model} relies only on data with responses observed (i.e., $R=1$), we can validate this model applying existing methods, such as the test based on unweighted sum of squares \citep{Hosmer1997},  available as \texttt{resid} in R package \texttt{rms}. 

For model \eqref{verification-model}, since the responses $Y$ are missed when $R=0$, we are unable to validate it directly.  However, based on the development in Section \ref{model-identification}, models \eqref{disease-model} and \eqref{verification-model} lead to model \eqref{drm.ch4}. Therefore, once model \eqref{disease-model} is validated, we can verify model \eqref{verification-model} by performing the goodness of fit test for model \eqref{drm.ch4}.
Motivated by the test based on the unweighted sum of squares for the standard logistic regression model, we can use
$$
T_2=\sum_{i=1}^n\left\{\left(R_i-\hat\pi_i\right)^2-\hat\pi_i\left\{1-\hat\pi_i\right)\right\}
$$
to construct a test statistic, 
where $\hat \pi_i=\pi(X_i,{\bm V}_i;\hat{\bm\mu},\hat {\bm \phi})$.
Let  ${se}(T_2)$ be the estimated standard error of $T_2$ obtained from the bootstrap method. 
Following \cite{Hosmer1997}, 
We use ${T}_2/ {se}(T_2)$ as the test statistic
and $N(0,1)$ as its reference distribution under the null hypothesis that model \eqref{verification-model} is correct. 

\section{Asymptotic properties}\label{chapter4-asymptotic}
 In this section, we establish the asymptotic distribution of $\widehat{AUC}$ and  $\widehat{ROC}(s)$, for each $s\in(0,1)$. 
We need some notation. 
Let $\bm{\theta}=(\bm{\mu}^T,\beta)^T$ and $\bm{\eta}=(\bm{\mu}^T,\beta,\psi_1,\psi_2,{\bm\psi}_3^T)^T$, 
whose dimensions are respectively $k_1$ and $k_2$.
We use $\bm{\theta}_0$ and $\bm{\eta}_0$
to denote the true values of $\bm{\theta}$ and $\bm{\eta}$. 
Write
$$
l_n(\bm\eta)=l_n({\bm \mu}, {\bm \phi})~~
\mbox{ and }~~
g(x,\bm{v} ,r;\bm{\theta} )
=g(x,\bm{v} ,r;\bm{\mu},\beta).
$$
Let $\lambda=P(Y=1)$, 
$\xi_{1-s}=F_0^{-1}(1-s)$, and 
$$
\bm{Z}_n=(\bm{Z}^T_{n1},Z_{n2},Z_{n3},Z_{n4},Z_{n5})^T,
$$
where
\begin{eqnarray*}
\bm{Z}_{n1}&=&n^{-1}
\frac{\partial l_n({\bm \eta}_0)}{\partial {\bm \eta}}, \\
Z_{n2}&=&n^{-1}\sum_{i=1}^n g_i\{F_0(X_i)-AUC\},\\
Z_{n3}&=&n^{-1}\sum_{i=1}^n \left(1-g_i\right) \left\{1-F_1(X_i)-AUC\right\},\\
Z_{n4}&=&n^{-1}\sum_{i=1}^n(1-g_i) \left\{I(X_i\leq\xi_{1-s})-F_0(\xi_{1-s})\right\},\\ 
Z_{n5}&=&n^{-1}\sum_{i=1}^ng_i\left\{I(X_i\leq\xi_{1-s})-F_1(\xi_{1-s})\right\},
\end{eqnarray*}
where $g_i=g(X_i,\bm{V}_i,R_i;\bm{\theta}_0)$. 
It can be verified that 
$$
E(\bm{Z}_{n})=
0.
$$
Denote the variance-covariance matrix of $\sqrt{n}\bm{Z}_n$ by ${\bm\Sigma}_Z$, and denote  the Fisher-information matrix at the true value ${\bm\eta}_0$ by 
$\bm{J}$. Let
$\bm{I}_{k_2,k_1}$ be a $k_2\times k_1$ matrix,
where the top $k_1\times k_1$ block is a identity matrix, and the rest elements are $0$'s. Define 
\begin{eqnarray*}
\bm{E}_1&=&E\left[\frac{\partial g(X,\bm{V},R;\bm{\theta}_0)}{\partial \bm{\theta}} \{F_0(X)-AUC\}\right],\\
\bm{E}_2&=&E\left[\frac{\partial g(X,\bm{V},R;\bm{\theta}_0)}{\partial \bm{\theta}}\left\{1-F_1(X)-AUC\right\}\right],\\
\bm{E}_3&=&E\left[\frac{\partial g(X,\bm{V},R;\bm{\theta}_0)}{\partial \bm{\theta}}\left\{I(X\leq\xi_{1-s})-F_0(\xi_{1-s})\right\}\right],\\
\bm{E}_4&=&E\left[\frac{\partial g(X,\bm{V},R;\bm{\theta}_0)}{\partial \bm{\theta}}\left\{I(X\leq\xi_{1-s})-F_1(\xi_{1-s})\right\}\right].
\end{eqnarray*}

The following theorem summarizes 
the asymptotic results of $\widehat{AUC}$ and $\widehat{ROC}(s)$.
The proof is given in Section 3 of the supplementary material. 

 \begin{theorem}
 \label{theorem.estimator}
Suppose Conditions C1--C6 in the Appendix are satisfied. As $n\to\infty$,  we have 
\begin{itemize}
\item[(a)] $\sqrt{n}(\widehat{AUC}-AUC)\overset{d}{\to}N(0,\sigma^2_{AUC})$, 
where $\sigma^2_{AUC}=\bm{H}_1 ^T\bm{\Sigma}_Z \bm{H}_1$ with $$\bm{H}_1 =\begin{pmatrix}
   & \bm{J}^{-1}\bm{I}_{k_2,k_1}\left(\frac{\bm{E}_1}{\lambda}-\frac{\bm{E}_2}{1-\lambda}\right)\vspace{0.05in}\\
   &\lambda^{-1}\vspace{0.05in}\\
   &(1-\lambda)^{-1}\vspace{0.05in}\\
   &0\vspace{0.05in}\\
 &  0
\end{pmatrix};$$
\item[(b)]
$
\sqrt{n}\left\{\widehat{ROC}(s)-ROC(s)\right\}\overset{d}{\to}N(0,\sigma^2_{s}),
$
where $\sigma^2_{s}=\bm{H}_2 ^T\bm{\Sigma}_Z \bm{H}_2$ with $$\bm{H}_2 =\begin{pmatrix}
   & \bm{J}^{-1}(\bm{\eta}_0)\bm{I}_{k_2,k_1}\left\{\frac{-\bm{E}_4}{\lambda}-\frac{\bm{E}_3}{1-\lambda}\frac{f_1(\xi_{1-s})}{f_0(\xi_{1-s})}\right\}\vspace{0.05in}\\
   &0\vspace{0.05in}\\
   &0\vspace{0.05in}\\
   &(1-\lambda)^{-1}\frac{f_1(\xi_{1-s})}{f_0(\xi_{1-s})}\vspace{0.05in}\\
   &-\lambda^{-1}\end{pmatrix},$$
   where $f_1(x)$ and $f_0(x)$ are the probability density functions of $F_1(x)$ and $F_0(x)$, respectively. 
\end{itemize}

\end{theorem}

The asymptotic variance formulas presented in Theorem \ref{theorem.estimator} can be used to construct plug-in variance estimators for $\widehat{AUC}$ and $\widehat{ROC}(s)$. The elements involved in the plug-in estimators take the form
$$
E\{h(X,{\bm V},R;\bm{\eta}_0,\lambda,F_0,F_1,f_0,f_1)\}.
$$
We can estimate $(\bm{\eta}_0,F_0,F_1)$ using $(\hat{\bm{\eta}},\hat F_0,\hat F_1)$. 
From \eqref{def.gfun}, we note that
$$
\lambda=P(Y=1)=E\{g(X,{\bm V},R;{\bm\theta}_0)\}.
$$
Thus, an estimator of $\lambda$ is given by 
$$
\hat\lambda=n^{-1}\sum_{i=1}^n \hat g_i.
$$
With $\hat F_0$ and $\hat F_1$, their corresponding densities
 $f_0$ and $f_1$ can be estimated by 
$$
\hat f_0(x)=\int K_{h_0}(u-x)d\hat F_0(u),~~
\hat f_1(x)=\int K_{h_1}(u-x)d\hat F_1(u),~~
$$
where $K_h(x)=K(x/h)/h$ with $K(\cdot)$ being a kernel function. 
Here, the bandwidths 
    $h_0$ and $h_1$ are chosen as 
$$h_0=1.06n^{-1/5}\min\left(\hat\sigma_0,\hat Q_0/1.34\right)~~
\mbox{and}  ~~
h_1=1.06n^{-1/5}\min\left(\hat\sigma_1,\hat Q_1/1.34\right),$$ 
where $\hat\sigma_0^2$ and $\hat \sigma_1^2$ are the variance estimators for the biomarker from the healthy and disease group; $\hat Q_0$ and $\hat Q_1$ are interquartile range estimators for the biomarker from the healthy and disease groups. 

With $(\hat{\bm{\eta}}_0,\hat \lambda,\hat F_0,\hat F_1,\hat f_0,\hat f_1)$, 
we can estimate
$
E\{h(X,{\bm V},R;\bm{\eta}_0,\lambda,F_0,F_1,f_0,f_1)\}
$
by
$$
n^{-1}\sum_{i=1}^n\{h(X_i,{\bm V}_i,R_i;\hat{\bm{\eta}}_0,\hat \lambda,\hat F_0,\hat F_1,\hat f_0,\hat f_1)\},
$$
which leads to the plug-in variance estimators 
$\hat\sigma_{AUC}^2$ for $\sigma^2_{AUC}$
and $\hat\sigma_{s}^2$ for $\sigma^2_{s}$ in Theorem \ref{theorem.estimator}. 
Consequently, the $100(1-\alpha)\%$ Wald-type confidence interval (CI) for $AUC$ is given by 
\begin{equation}
\label{AUC.ci.ch4}
[\widehat{AUC}-z_{1-\alpha/2}\hat\sigma_{AUC}/\sqrt{n}, \quad  
\widehat{AUC}+z_{1-\alpha/2}\hat\sigma_{AUC}/\sqrt{n}]
,
\end{equation}
and 
likewise, that for $ROC(s)$ is given by 
\begin{equation}
\label{ROC.ci.ch4}
[\widehat{ROC}(s)-z_{1-\alpha/2}\hat\sigma_{s}/\sqrt{n}, \quad 
\widehat{ROC}(s)+z_{1-\alpha/2}\hat\sigma_{s}/\sqrt{n}].
\end{equation}

\section{Simulation studies}
\label{chapter4-simulation}

In this section, we compare the performance of our method with existing methods in the AUC and ROC curve estimation through simulation examples. We also evaluate the accuracy of the Wald-type CIs in \eqref{AUC.ci.ch4} and \eqref{ROC.ci.ch4} at the nominal level $1-\alpha = 0.95$.

\subsection{Simulation setup}\label{simulation-setup}
Our method (denoted as ``Our") is compared with four competitive methods: 
\begin{itemize}
 \item[--]  the IPW method in \cite{Yu2018}, denoted as ``IPW";
 \item[--]  the  method that  assumes an ignorable mechanism in the verification model, i.e., $\beta=0$ in \eqref{verification-model}, denoted as ``IG";
 \item[--]  the method based only on the verified data (i.e. $R=1$), denoted as ``VER";
 \item[--]  the full data method, which assumes all the data are available, denoted as ``Full".
  This method serves as a benchmark.
\end{itemize}

 We do not include the method of \cite{Liu2010} as a competitive method as their model assumptions cannot be verified using the observed data. For the data simulation, 
the biomarker $X$ is generated from $\text{Uniform}[-1,1]$.
The covariates are simulated as two-dimensional: ${\bm V}=(V_1,V_2)^T$, where $V_1\sim N(0,1)$ and $V_2\sim \text{Bernoulli}(0.5)$.
Here, $X$, $V_1$, and $V_2$ are simulated as independent. 
To simulate $Y$ and $R$, we consider the following three scenarios. 

{\it Scenario 1}:  
\begin{eqnarray*}
    P(Y=1|X=x,{\bm V}={\bm v},R=1)&=&\frac{1}{1+\exp\left(1.7-2.5x-1.5v_1-1.5 v_2\right)},\\
    P(R=1|X=x,{\bm V}={\bm v},Y=y)&=&\frac{1}{1+\exp\left(1.3-1.5x-1.2v_1+v_2\right)}.
\end{eqnarray*}
 
 {\it Scenario 2}:  
\begin{eqnarray*}
    P(Y=1|X=x,{\bm V}={\bm v},R=1)&=&\frac{1}{1+\exp\left(1.7-2.5x-1.5v_1-1.5 v_2\right)},\\
    P(R=1|X=x,{\bm V}={\bm v},Y=y)&=&\frac{1}{1+\exp\left(1.3-1.5x-1.2v_1+v_2-2y\right)}.
\end{eqnarray*}

{\it Scenario 3}:  
\begin{eqnarray*}
    P(Y=1|X=x,{\bm V}={\bm v},R=1)&=&\frac{1}{1+\exp\left(1.7-2.5x-1.5v_1+0.5v_1^2-1.5 v_2\right)},\\
    P(R=1|X=x,{\bm V}={\bm v},Y=y)&=&\frac{1}{1+\exp\left(1.3-1.5x-1.2v_1+0.5v_1^2+v_2-2y\right)}.
\end{eqnarray*}


Scenario 1 corresponds to the case where the missing mechanism is MAR; Scenario 2 corresponds to the case of a non-ignorable missing mechanism. In both scenarios, the models of our method and the IPW method are correctly specified.
In Scenario $3$, a second order term $v_1^2$ is added to both the verification and disease models, resulting in model misspecification for both our and the IPW method. 

Summary quantities for these scenarios are displayed in Table \ref{simu-quantity}. 
 For each scenario, we consider sample sizes: $n= 5000$ and $ n = 10000$. 
 For each simulation setting, we repeat 1000 times.

\begin{table}[!htbp]
\caption{Summary quantities for three scenarios}\label{simu-quantity}
\centering
\begin{tabular*}{\textwidth}{c@{\extracolsep{\fill}} c c c c } 
\toprule
 &\multicolumn{1}{c}{Scenario 1}&\multicolumn{1}{c}{Scenario 2}&\multicolumn{1}{c}{Scenario 3} \\ 
 \cline{2-2}\cline{3-3}\cline{4-4}
P(Y=1) & 0.369 & 0.245&0.191\\
P(R=1) & 0.219 & 0.307 & 0.242 \\
AUC & 0.751 & 0.776 & 0.813 \\
ROC(0.1) & 0.347 & 0.369 & 0.418 \\
ROC(0.2) & 0.548 & 0.587 & 0.657 \\
\bottomrule
\end{tabular*}
\end{table}

\subsection{AUC point estimation results}
\label{simulation-AUC}
In this section, we compare our method with the competitive methods in the AUC estimation. 
The criteria for comparison are the relative bias (RB) and mean square error (MSE). In particular, for AUC estimates from a method: $\hat a^{(i)}, i = 1, \ldots, B$, the corresponding RB (in percentage) and MSE are defined to be:
\begin{equation*}
RB(\%)=\frac{1}{B}\sum_{i=1}^B \frac{\hat a^{(i)}-a_0}{a_0}\times100~~{\rm and}~~MSE=\frac{1}{B}\sum_{i=1}^B (\hat a^{(i)}-a_0)^2,
\end{equation*}
where $a_0$ is the true value of the AUC. 
The results are summarized in  Table \ref{AUC}.

\begin{table}[!htbp]
\caption{RB ($\%$) and MSE ($\times 1000$) of AUC estimates }\label{AUC}
\centering
\begin{tabular*}{\textwidth}{c@{\extracolsep{\fill}} c c c c c c c c c c c} 
\toprule
$n$& Method &\multicolumn{2}{c}{Scenario 1}&\multicolumn{2}{c}{Scenario 2}&\multicolumn{2}{c}{Scenario 3} \\ 
 \cline{3-4}\cline{5-6}\cline{7-8}
& & Rb & MSE&Rb & MSE&Rb & MSE\\
\hline
5000&Our&-0.079&0.213&-0.129&0.163&-0.221&0.213\\
5000&IPW&-0.629&0.902&-0.216&0.352&0.720&0.299\\
5000&IG&-0.001&0.173&-3.253&0.775&-4.524&1.555\\
5000&VER&-5.597&2.013&-10.740&7.142&-9.809&6.561\\
5000&Full&-0.020&0.048&-0.017&0.055&-0.013&0.049\\
\hline
10000&Our&-0.107&0.117&-0.025&0.059&-0.199&0.104\\
10000&IPW&-0.222&0.384&-0.034&0.141&0.826&0.157\\
10000&IG&-0.060&0.089&-3.233& 0.698&-4.526&1.457\\
10000&VER&-5.616&1.910&-10.698&6.989&-9.796&6.452\\
10000&Full&-0.024&0.025&0.034&0.026&0.034&0.025\\
\bottomrule
\end{tabular*}
\end{table}

From the table, we observe that the Full method results in the smallest MSEs across all simulation setups, as expected. Therefore, we exclude the Full method from further discussion. The VER method exhibits the largest absolute RBs compared to the other methods in all scenarios. When compared to the IPW method, our method consistently achieves much smaller MSEs in every scenario. In Scenario 1, where the missingness mechanism for disease status is MAR, the IG method yields the smallest MSE. In this case, our proposed method performs similarly to the IG method, with no significant difference. However, in the other scenarios, our proposed method shows a clear advantage over the IG method in terms of MSE.

\subsection{AUC CIs estimation results}
\label{simulation-CI}

In this section, we evaluate the performance of confidence interval (CI) estimation for the proposed method at the nominal level of 0.95. Two measures are used to assess the effectiveness of the CI estimation: coverage probability (CP) and average length (AL).
Let $\widehat{\text{CI}}^{(i)}$ denote the $i$-th estimated CI for the AUC, where $i = 1, \dots, B$.  The CP and  AL are defined to be:
\begin{equation*}
CP=\frac{1}{B}\sum_{i=1}^B I(a_0\in \widehat {CI}^{(i)})~~{\rm and}~~AL=\frac{1}{B}\sum_{i=1}^B |\widehat {CI}^{(i)}|,
\end{equation*}
where $a_0$ is the true value of the AUC. 
We also compute the CPs and ALs of CIs based on the IPW method for comparison.
The results are shown in Table \ref{CI}.

\begin{table}[!htbp]
\caption{CP  and AL  of CIs for AUC by IPW and Our methods}\label{CI}
\centering
\begin{tabular*}{\textwidth}{c@{\extracolsep{\fill}} c c c c c c c c} 
\toprule
$n$& Method&\multicolumn{2}{c}{Scenario 1}&\multicolumn{2}{c}{Scenario 2}&\multicolumn{2}{c}{Scenario 3} \\ 
 \cline{3-4}\cline{5-6}\cline{7-8}
&&CP & AL&CP & AL&CP & AL&\\
\hline
5000&Our&0.949 &0.058&0.944 &0.038&0.927 &0.050\\
5000&IPW&0.948& 0.110&0.976& 0.084&0.914 &0.070\\
\hline
10000&Our&0.940 &0.041&0.956 &0.026&0.935 &0.035\\
10000&IPW&0.941& 0.075&0.980& 0.054&0.869 &0.045\\
\bottomrule
\end{tabular*}
\end{table}

We first focus on the results for coverage probability (CP).
In cases where the models are correctly specified (Scenarios 1 and 2), the CPs for our proposed method are reasonably close to the nominal level. In Scenario 1, the IPW method performs similarly to our proposed method. However, in Scenario 2, the IPW method exhibits an overcoverage problem.
In the case of model misspecification (Scenario 3), both methods show an undercoverage problem. The CPs for our proposed method are not significantly smaller than the nominal level, whereas the undercoverage problem is more pronounced for the IPW method.
Regarding average length (AL), our proposed method consistently yields a significantly smaller AL across all scenarios compared to the IPW method.

\subsection{ROC curve point estimation results}
\label{simulation-ROC}

In this section, we compare our method with competitive methods for ROC curve estimation. Since \cite{Yu2018} does not discuss the estimation of the ROC curve, their method is not included in the comparison.

To evaluate ROC curve estimation, we estimate the ROC curve at $s = 0.1$ and $s = 0.2$, corresponding to specificities of 0.9 and 0.8, respectively. Similar to Section \ref{simulation-AUC}, the criteria for comparison are RB and MSE. The simulation results for $ROC(0.1)$ and $ROC(0.2)$ are summarized in Table \ref{ROC0.9}.

Similar to the AUC estimation discussed in Section \ref{simulation-AUC}, for both $ROC(0.1)$ and $ROC(0.2)$, the Full method yields the smallest MSE, while the VER method exhibits the largest MSE among all methods across all scenarios. 

When the missingness mechanism for disease status is MAR (Scenario 1), the IG method and our proposed method show similar MSEs.  However, in other scenarios, our proposed method demonstrates a smaller MSE, with the difference being significant.

\begin{table}[!htbp]
\caption{RB ($\%$) and MSE ($\times 1000$) of $ROC(s)$ for $s=0.1$ and $0.2$}\label{ROC0.9}
\centering
\begin{tabular*}{\textwidth}{c@{\extracolsep{\fill}} c c c c c c c c c c c c} 
\toprule
$s$&$n$& Method &\multicolumn{2}{c}{Scenario 1}&\multicolumn{2}{c}{Scenario 2}&\multicolumn{2}{c}{Scenario 3} \\ 
 \cline{4-5}\cline{6-7}\cline{8-9}
&&& RB & MSE&RB & MSE&RB & MSE\\
\hline
0.1&5000&Our&0.306&0.524&-0.095&0.446&3.553&0.889\\
0.1&5000&IG&0.564&0.523&-5.566&0.838&-6.629&1.428\\
0.1&5000&VER&-23.116&7.705&-34.606&17.178&-33.820&21.260\\
0.1&5000&Full&0.474&0.314&0.015&0.362&0.074&0.416\\
\hline
0.1&10000&Our&0.134&0.283&0.152&0.198&3.644&0.589\\
0.1&10000&IG&0.306&0.269&-5.540&0.631&-6.656&1.129\\
0.1&10000&VER&-23.439&7.304&-34.452&16.628&-33.953&20.823\\
0.1&10000&Full&0.376&0.167&0.314&0.187&0.361&0.246\\
\midrule
0.2&5000&Our&0.192&0.644&-0.528&0.564&-0.039&0.834\\
0.2&5000&IG&0.343&0.596&-6.390&1.899&-9.279&4.452\\
0.2&5000&VER&-16.351&9.370&-27.409&26.895&27.009&32.718\\
0.2&5000&Full&0.296&0.278&-0.320&0.337&-0.472&0.378\\
\hline
0.2&10000&Our&0.066&0.357&-0.342&0.230&-0.034&0.406\\
0.2&10000&IG&0.172&0.314&-6.377&1.647&-9.316&4.117\\
0.2&10000&VER&-16.295&8.714&-27.214&26.020&26.811&31.746\\
0.2&10000&Full&0.261&0.137&-0.223&0.154&-0.399&0.185\\
\bottomrule
\end{tabular*}
\end{table}

\subsection{ROC curve CIs estimation results}
\label{simulation-ROC-CI}

In this section, we assess the performance of the CI estimation of the ROC curve using the proposed  method at the nominal level of $0.95$.
Specifically, we estimate the CIs for $ROC(0.1)$ and $ROC(0.2)$.
Similar to Section \ref{simulation-CI}, we quantitatively evaluate the effectiveness of the CI estimation using  CP and AL.
The results are presented in Table \ref{ROC-CI}.

\begin{table}[!htb]
\caption{CP  and AL of the estimated CIs for $ROC(s)$ with $s=0.1$ and $s=0.2$}\label{ROC-CI}
\centering
\begin{tabular*}{\textwidth}{c@{\extracolsep{\fill}} c c c c c c c c} 
\toprule
$s$&$n$ &\multicolumn{2}{c}{Scenario 1}&\multicolumn{2}{c}{Scenario 2}&\multicolumn{2}{c}{Scenario 3} \\ 
 \cline{3-4}\cline{5-6}\cline{7-8}
&  &CP & AL&CP & AL&CP & AL&\\
\hline
0.1& 5000&0.952& 0.092&0.939& 0.074&0.898 &0.097\\
0.1&10000&0.943 &0.064&0.953 &0.052&0.837 &0.068\\
0.2&5000&0.960& 0.103&0.943& 0.077&0.922 &0.102\\
0.2&10000&0.950 &0.073&0.954 &0.054&0.931 &0.071\\
\bottomrule
\end{tabular*}
\end{table}

Under Scenarios 1 and 2, the CPs for our proposed method are reasonably close to the nominal level for both $s=0.1$ and $0.2$.
When the models are misspecified (Scenario 3), the CPs fall below the nominal level  for both $s=0.1$ and $0.2$.
In this scenario, the undercoverage problem is more pronounced for $s=0.1$.

\section{Real data application}
\label{realdata}

In this section, we apply our proposed method to data collected by the National Alzheimer’s Coordinating Center (NACC). The dataset analyzed is the Uniform Data Set (UDS), which includes all visits for all participants as of the December 2023 data freeze, comprising 185,831 patients. Our focus here is on evaluating the diagnostic ability of the Mini Mental State Examination (MMSE) in detecting Alzheimer's disease (AD).
The MMSE, originally created by \cite{folstein1975}, has since been shown to be effective in quantitatively assessing the severity of cognitive impairment \citep{Tombaugh1992}. It consists of a series of questions, with scores ranging from 0 to 30, where lower scores indicate greater cognitive impairment.
For this analysis, we use the NIA-AA Alzheimer’s disease neuropathologic change (ADNC) (ABC score) as the gold standard. Since the ABC score requires a brain autopsy, some patients or their families may decline the procedure, leading to missing disease status data. The test results are categorized as ``Not AD", ``Low ADNC", ``Intermediate ADNC", and ``High ADNC". For our study, we define ``Not AD" as the healthy group and group the remaining categories under ``AD".

Since missing disease status is strongly correlated with mortality, and individuals who are alive are generally assumed to be in better health, the verification of disease status is likely to be nonignorable. 
Following \cite{Liu2010}, we consider age, sex, race, marital status, stroke, Parkinson's disease, and depression as relevant covariates.
For illustration, we focus on patients aged 65 to 70. Patients with missing MMSE results or covariates are excluded from the analysis, resulting in a total of 15,564 patients. Table \ref{nacc-summary} provides a summary of the MMSE scores and relevant covariates for this subgroup.

\begin{table}[!htbp]
\caption{Summary statistics for NACC data on patients aged 65 to 70}\label{nacc-summary}
\centering
\begin{tabular*}{\textwidth}{l@{\extracolsep{\fill}} l l} 
\toprule
Variable & &Summary statistics\\
\hline
$Y$ & &AD: 9.0$\%$; AD-free: 1.2$\%$; missing: 89.8$\%$ \\
MMSE & & Mean: 26.3; SD: 5.8; Median: 29\\
Age & & Mean: 67.7; SD: 1.7; Median: 68\\
Sex & & Male: 39.6$\%$; Female: 60.4$\%$\\
Race& & White: 82.5$\%$; other: 17.5$\%$\\
Martial status& &  Married: 69.4$\%$; other: 30.6$\%$\\
Stroke & & Yes: 2.0$\%$; No: 98.0$\%$\\
Parkinson's disease & & Yes: 2.6$\%$; No: 97.4$\%$\\
Depression & & Yes: 19.4$\%$; No: 80.6$\%$\\
\bottomrule
\end{tabular*}
\end{table}

In our data analysis, we define the biomarker as  $X=(30 - MMSE) / 15$ and transform age to $(age - 65) / 5$ for ease of coefficient reporting. Additionally, we denote the relevant covariates by ${\bm V}$. We consider the following disease and verification models:
\begin{eqnarray} \label{nacc-disease}
P(Y=1|X=x,{\bm V}={\bm v},R=1)=\frac{1}{1+\exp(\mu_1+\mu_2x+{\bm\mu}_3^T{\bm v})},   
\end{eqnarray}
and  
\begin{eqnarray}\label{nacc-verification}
P(R=1|X=x,{\bm V}={\bm v},Y=y)=\frac{1}{1+\exp(\psi_1+\psi_2 x+{\bm\psi}_3^T{\bm v}+\beta y)}.
\end{eqnarray}
The two-step procedure outlined in Section \ref{model verification} yields a p-value of 0.557 for the goodness of fit of model \eqref{nacc-disease} and a p-value of 0.194 for the goodness of fit of model \eqref{nacc-verification}. These results indicate that models \eqref{nacc-disease} and \eqref{nacc-verification} collectively provide a reasonable fit to the data.

Next, we apply our proposed method  using models \eqref{nacc-disease} and \eqref{nacc-verification}. The estimated model parameters and their standard errors are summarized in Table \ref{nacc-parameter}.
In the verification model, we find that the non-ignorability parameter 
$\beta$ is highly significant, confirming that the missing data mechanism is non-ignorable. Other significant covariates include age, sex, race, and marital status. In the disease model, MMSE, age, and depression are significant at the 5\% level.
Since our primary focus is on estimating the AUC and ROC curve, and given the relatively small estimated coefficients for the other covariates, we retain the full model without conducting further variable selection in the subsequent analysis.

\begin{table}[!htbp]
\caption{Estimated model parameters and their standard errors (Significant coefficients at the 5\% level are bolded) }\label{nacc-parameter}
\centering
\begin{tabular*}{\textwidth}{l@{\extracolsep{\fill}} r r} 
\toprule
Variable & Verification model & Disease model\\
\hline
Intercept &\textbf{5.379 (0.219)}&\textbf{-1.216 (0.277)}\\
(30-MMSE)/15 & -0.237 (0.391)&\textbf{-2.692 (0.139)}\\
(Age-65)/5 &\textbf{0.303 (0.153)}&\textbf{-0.567 (0.196)}\\
Sex &\textbf{-0.384 (0.103)}& -0.266 (0.150)\\
Race& \textbf{-1.239 (0.178)}& 0.248 (0.255)\\
Martial status&\textbf{-0.276 (0.132)} & 0.035 (0.197)\\
Stroke &-0.070 (0.305) & -0.194 (0.492)\\
Parkinson's disease & -0.002 (0.246)&-0.166 (0.355)\\
Depression & 0.070 (0.117)& \textbf{-0.363 (0.168)}\\
Y (gold standard) &\textbf{-3.099 (0.794)} & \\
\bottomrule
\end{tabular*}
\end{table}

\begin{figure}[!ht]
 \begin{center}
 \includegraphics[width=14cm, height=14cm]{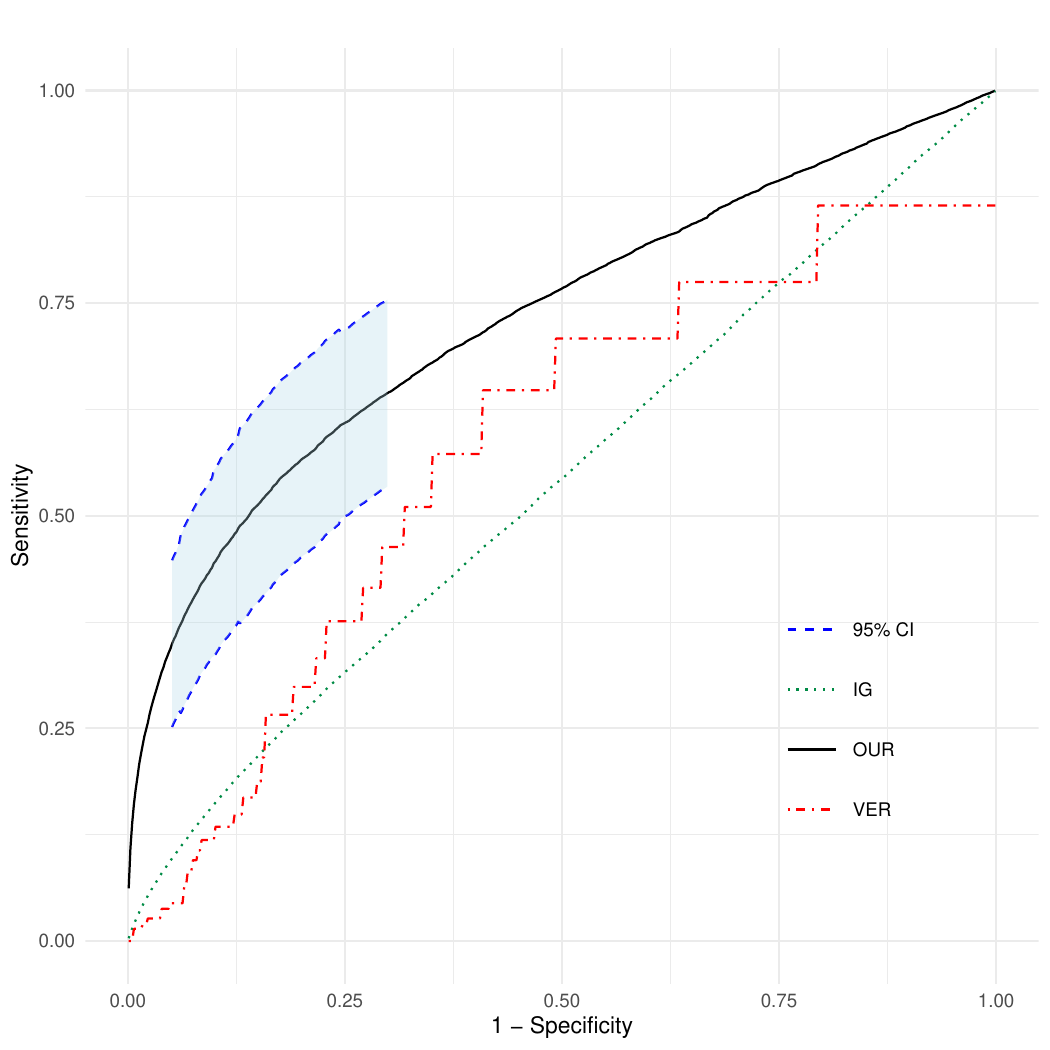}
\end{center}
\caption{
Estimated ROC curves and  $95\%$ confidence band for $s\in [0.05,0.3]$ 
}
\label{Figure: ROC curve}
\end{figure}

In terms of AUC estimation, we apply the methods introduced in Section \ref{simulation-setup} to the NACC data. We estimate the CI of the AUC using both the IPW method and our proposed method.
The AUC and its CI estimates are presented in Table \ref{nacc-auc}. Our proposed method yields the largest AUC with a shorter CI estimate compared to the IPW method.
Figure \ref{Figure: ROC curve} displays the estimated ROC curve using the IG, VER, and our proposed method.
For our proposed method, we also present the $95\%$ confidence band (the shaded area) for $s\in (0.05, 0.3)$.
Our proposed method demonstrates a higher sensitivity value at each specificity level compared to the IG and VER methods.

\begin{table}[!htb]
\caption{AUC and 95$\%$ CI estimates for the NACC data}\label{nacc-auc}
\centering
\begin{tabular*}{\textwidth}{l@{\extracolsep{\fill}} c c r} 
\toprule
Methods & AUC & CI & Length\\
\hline
Our & 0.791 & (0.779, 0.803)& 0.024\\
IPW & 0.744 & (-1.289, 2.778)& 4.067\\
IG & 0.584 & ------ & ------\\
VER & 0.614 &------ &------ \\
\bottomrule
\end{tabular*}
\end{table}

\section{Conclusion}
\label{conclude}

In this article, we propose a likelihood-based approach to estimate the ROC curve and the AUC when the missing mechanism of disease status is non-ignorable. We establish the identifiability of the model parameters under mild conditions and demonstrate the asymptotic normality of the proposed estimators for both the ROC curve and the AUC. Simulation studies illustrate the advantages of our method compared to existing approaches. Finally, we apply our proposed method to the Alzheimer's disease dataset from NACC to analyze the diagnostic accuracy of the MMSE test.

In this paper, we focus on binary responses. However, ordinary responses with multiple categories may arise in applications, such as those found in the NACC data. We plan to extend the proposed method to accommodate these situations and will explore this in future research.


\section*{Acknowledgments}
Dr. Li's research is supported in part by the Natural Sciences and Engineering Research Council of Canada. Dr. Yu’s research is supported in part by the Singapore Ministry of Education Academic Research Tier 1 Fund: A-8000413-00-00.
The NACC database is funded by NIA/NIH Grant U24 AG072122. NACC data are contributed by the NIA-funded ADRCs: P30 AG062429 (PI James Brewer, MD, PhD), P30 AG066468 (PI Oscar Lopez, MD), P30 AG062421 (PI Bradley Hyman, MD, PhD), P30 AG066509 (PI Thomas Grabowski, MD), P30 AG066514 (PI Mary Sano, PhD), P30 AG066530 (PI Helena Chui, MD), P30 AG066507 (PI Marilyn Albert, PhD), P30 AG066444 (PI David Holtzman, MD), P30 AG066518 (PI Lisa Silbert, MD, MCR), P30 AG066512 (PI Thomas Wisniewski, MD), P30 AG066462 (PI Scott Small, MD), P30 AG072979 (PI David Wolk, MD), P30 AG072972 (PI Charles DeCarli, MD), P30 AG072976 (PI Andrew Saykin, PsyD), P30 AG072975 (PI Julie A. Schneider, MD, MS), P30 AG072978 (PI Ann McKee, MD), P30 AG072977 (PI Robert Vassar, PhD), P30 AG066519 (PI Frank LaFerla, PhD), P30 AG062677 (PI Ronald Petersen, MD, PhD), P30 AG079280 (PI Jessica Langbaum, PhD), P30 AG062422 (PI Gil Rabinovici, MD), P30 AG066511 (PI Allan Levey, MD, PhD), P30 AG072946 (PI Linda Van Eldik, PhD), P30 AG062715 (PI Sanjay Asthana, MD, FRCP), P30 AG072973 (PI Russell Swerdlow, MD), P30 AG066506 (PI Glenn Smith, PhD, ABPP), P30 AG066508 (PI Stephen Strittmatter, MD, PhD), P30 AG066515 (PI Victor Henderson, MD, MS), P30 AG072947 (PI Suzanne Craft, PhD), P30 AG072931 (PI Henry Paulson, MD, PhD), P30 AG066546 (PI Sudha Seshadri, MD), P30 AG086401 (PI Erik Roberson, MD, PhD), P30 AG086404 (PI Gary Rosenberg, MD), P20 AG068082 (PI Angela Jefferson, PhD), P30 AG072958 (PI Heather Whitson, MD), P30 AG072959 (PI James Leverenz, MD).

\section*{Appendix: Regularity conditions}
Let ${\bm Z}=(X,{\bm V}^T,Y,R)^T$, 
${\bm z}=(x,{\bm v}^T,y,r)^T$, 
and 
\begin{align*}
l({\bm z}; {\bm \eta})=&r\left[y\log \left\{p_{1}(x,\bm{v};\bm{\mu})\right\}+(1-y)\log \left\{1-p_{1}(x,\bm{v};\bm{\mu})\right\}\right]\\ &+\left [r\log \{ \pi(x,{\bm v};{\bm\mu},{\bm \phi})\} +(1-r) \log\{1- \pi(x,{\bm v};{\bm\mu},{\bm \phi}) \}\right].
\end{align*}
The results in Theorems \ref{theorem-identifiability} and \ref{theorem.estimator} rely on the following regularity conditions. 
\begin{enumerate}
    \item[C1.] The biomarker $X$ is continuous, and the components of $(1, X, \mathbf{V})$ are stochastically linearly independent. Furthermore, the coefficient $\mu_2$ in model \eqref{disease-model} is non-zero. 
    
     \item[C2.] The parameter space ${\bm \Theta}$ of ${\bm \eta}$
     is compact and 
     the true value ${\bm \eta}_0$
      of ${\bm \eta}$ is the interior point of the parameter space  ${\bm \Theta}$.

    \item[C3.] 
    (i) $E(|\log l({\bm Z}; {\bm \eta}_0)| )<\infty$;
(ii) for sufficiently small $\rho$ and for sufficient large $t$,
$E[ \log (1+ \log l({\bm Z}; {\bm \eta}, \rho) ]<\infty$ for ${\bm\eta} \in {\bm \Theta}$
and
$E[ \log(1+ \varphi({\bm Z}; t) ]<\infty$,
where 
$$
l({\bm Z}; {\bm \eta}, \rho)= \sup _{\|{\bm \eta}' -{\bm \eta}\|<\rho} l({\bm Z}; {\bm \eta}')~~
\mbox{
and 
}
~~
\varphi({\bm Z}; t) =\sup _{\| {\bm \eta} \| \geq t} l({\bm Z}; {\bm \eta});
$$
(iii) $l({\bm Z}; {\bm \eta}) \to 0$ in probability as $\|{\bm \eta}\| \to \infty$.
Here $\|{\bm \eta}\|$ is the $L_2$-norm of ${\bm \eta}$.

    \item[C4.] The Fisher information matrix ${\bm J}$ exists and is positive definite.
    \item[C5.] Let $\eta_1$, $\eta_2$, and $\eta_3$ be any three elements of ${\bm \eta}$. 
    There exists a positive $\epsilon$ 
    and a function $H$ such that 
    $$
\sup_{\|{\bm \eta}-{\bm \eta}_0\|\leq \epsilon}\left |
\frac{\partial l^3({\bm Z}; {\bm \eta})}
{\partial\eta_1\partial\eta_2\partial\eta_3}
\right |\leq H({\bm Z}),$$
where $E\{H({\bm Z})\}<\infty$. 

    \item[C6.]
        The first and second moments of the biomarker $X$ and covariates ${\bm V}$ exist and are finite.
    
\end{enumerate}

\bibliographystyle{apalike} 

\bibliography{Missing}

 \end{document}